\documentclass[conference]{IEEEtran}
\IEEEoverridecommandlockouts

\usepackage{cite}
\usepackage{amsmath,amssymb,amsfonts}
\usepackage{algorithmic}
\usepackage{graphicx}
\usepackage{textcomp}
\usepackage{booktabs}
\usepackage{array}
\usepackage{multirow}
\usepackage{url}
\usepackage{hyperref}
\usepackage[table]{xcolor}
\usepackage{threeparttable}
\usepackage{listings}
\def\BibTeX{{\rm B\kern-.05em{\sc i\kern-.025em b}\kern-.08em
    T\kern-.1667em\lower.7ex\hbox{E}\kern-.125emX}}

\title{When Agents Fail to Act: A Diagnostic Framework for Tool Invocation Reliability in Multi-Agent LLM Systems}

\newif\ifshowchanges
\showchangesfalse

\ifshowchanges
    \newcommand{\add}[1]{\textcolor{blue}{#1}}
    \newcommand{\delete}[1]{\textcolor{red}{\st{#1}}}
    
\else
    \newcommand{\add}[1]{#1}
    \newcommand{\delete}[1]{}
    
\fi

\begin{document}

\title{LLM-as-a-Judge for Scalable Test Coverage Evaluation: Accuracy, Operational Reliability, \\and Cost}

% \author{
%     Donghao Huang\textsuperscript{\rm 1,2},
%     Shila Chew\textsuperscript{\rm 3},
%     Anna Dutkiewicz\textsuperscript{\rm 2},
%     Zhaoxia Wang\textsuperscript{\rm 1}
% }

\author{
\IEEEauthorblockN{Donghao Huang\textsuperscript{\rm 1,2},
Shila Chew\textsuperscript{\rm 3},
Anna Dutkiewicz\textsuperscript{\rm 2},
Zhaoxia Wang\textsuperscript{\rm 1}}
\IEEEauthorblockA{\textsuperscript{1}School of Computing and Information Systems, Singapore Management University, Singapore}
\IEEEauthorblockA{\textsuperscript{2}Research and Development, Mastercard, Arlington, VA, USA}
\IEEEauthorblockA{\textsuperscript{2}Research and Development, Mastercard, Singapore, Singapore}
\IEEEauthorblockA{dh.huang.2023@smu.edu.sg, shila.chew@mastercard.com, anna.dutkiewicz@mastercard.com, zxwang@smu.edu.sg}
}

% \affiliations{
%     \textsuperscript{\rm 1}School of Computing and Information Systems, Singapore Management University, Singapore\\
%     \textsuperscript{\rm 2}Research and Development, Mastercard, Arlington, VA, USA\\
%     \textsuperscript{\rm 3}Research and Development, Mastercard, Singapore, Singapore\\
%     dh.huang.2023@smu.edu.sg, shila.chew@mastercard.com, anna.dutkiewicz@mastercard.com, zxwang@smu.edu.sg
% }

\maketitle

\begin{abstract}
Assessing software test coverage at scale remains a bottleneck in QA pipelines. We present \emph{LLM-as-a-Judge (LAJ)}, a production-ready, rubric-driven framework for evaluating Gherkin acceptance tests with structured JSON outputs. Across 20 model configurations (GPT-4, GPT-5 with varying reasoning effort, and open-weight models) on 100 expert-annotated scripts over 5 runs (500 evaluations), we provide the first comprehensive analysis spanning accuracy, operational reliability, and cost. We introduce the \emph{Evaluation Completion Rate} (ECR@1) to quantify first-attempt success, revealing reliability from 85.4\% to 100.0\% with material cost implications via retries. Results show that smaller models can outperform larger ones: GPT-4o Mini attains the best accuracy (6.07 MAAE), high reliability (96.6\% ECR@1), and low cost (\$1.01 per 1K), yielding a 78$\times$ cost reduction vs.\ GPT-5 (high reasoning) while improving accuracy. Reasoning effort is model-family dependent: GPT-5 benefits from increased reasoning (with predictable accuracy--cost trade-offs), whereas open-weight models degrade across all dimensions as reasoning increases. Overall, cost spans 175$\times$ (\$0.45--\$78.96 per 1K). We release the dataset, framework, and code to support reproducibility and deployment.
\end{abstract}

\section{Introduction}

Automated assessment of software test coverage presents a persistent challenge in quality assurance: manual evaluation by domain experts is accurate but prohibitively expensive and slow at the scales required for modern continuous integration/continuous deployment (CI/CD) pipelines. Traditional static analysis tools provide quantitative metrics (e.g., line coverage, branch coverage) but cannot assess the \textit{semantic completeness} of test scenarios---whether tests adequately capture business requirements, edge cases, and error conditions~\cite{karpurapu2024comprehensive}.

Recent advances in large language models (LLMs) have demonstrated capabilities in code understanding, semantic analysis, and structured evaluation tasks~\cite{alshahwan2024automated}. This raises a compelling question: can LLMs serve as reliable, scalable, and cost-effective judges for evaluating test coverage quality? While the concept of "LLM-as-a-Judge" has gained traction in various evaluation contexts~\cite{zheng2023judging}, systematic investigation of its application to software testing---particularly with respect to accuracy, cost, and \textit{operational reliability}---remains limited.

In this work, we introduce a comprehensive \textbf{LLM-as-a-Judge (LAJ)} framework specifically designed for evaluating Gherkin-style acceptance test coverage\footnote{Gherkin is a domain-specific language for behavior-driven development (BDD) that uses Given-When-Then syntax to specify test scenarios in natural language.}. Our framework employs rubric-driven assessment aligned with industry best practices and HTTP method-specific testing guidelines. Beyond measuring assessment accuracy, we introduce novel metrics that capture operational reliability: the \textit{Evaluation Completion Rate (ECR@1)}, which measures the percentage of evaluations that succeed on the first attempt, and \textit{adjusted cost metrics} that account for retry overhead in production deployments.

Our main contributions are as follows:
\begin{itemize}
    \item We introduce \emph{LLM-as-a-Judge (LAJ)}, a production-ready framework for automated test coverage evaluation that uses rubric-driven assessment.% and structured output validation.
    \item We propose reliability-aware metrics—\emph{Evaluation Completion Rate (ECR@1)} and reliability-adjusted cost—that capture deployment-critical robustness beyond accuracy alone.
    \item We conduct a comprehensive evaluation of 20 model configurations across 500 runs, % (100 test scripts $\times$ 5 repetitions), 
    systematically quantifying accuracy–reliability–cost trade-offs spanning a 175$\times$ range (\$0.45--\$78.96 per 1K evaluations).
    \item We provide evidence-based deployment recommendations, identifying GPT-4o Mini as production-optimal (6.07 MAAE, 96.6\% ECR@1, \$1.01/1K), and publicly release our dataset, evaluation framework, and implementation for reproducibility.\add{\footnote{GitHub repository: \url{https://github.com/inflaton/LAJ-Gherkin}.}}
\end{itemize}

% \paragraph{Contributions.}
% (i) We introduce \emph{LLM-as-a-Judge (LAJ)}, a production-ready, rubric-driven framework for automated test coverage evaluation with standardized JSON outputs. 
% (ii) We propose reliability-aware metrics—\emph{Evaluation Completion Rate (ECR@1)} and adjusted cost—that capture deployment-critical reliability, revealing operational performance ranging from 85.4\% to 100.0\%. 
% (iii) We conduct a multi-dimensional evaluation of 20 model configurations across 500 runs (100 scripts $\times$ 5), quantifying accuracy–reliability–cost trade-offs spanning 175$\times$ (\$0.45--\$78.96 per 1K). 
% (iv) We provide evidence-based deployment guidance, identifying GPT-4o Mini as production-optimal (6.07 MAAE, 96.6\% ECR@1, \$1.01/1K), and publicly release the dataset, framework, and code for reproducibility.\footnote{GitHub repository link to be released upon paper acceptance.}

\section{Related Work}

\add{\textbf{Traditional Test Coverage Tools.} Established static analysis tools like JaCoCo~\cite{crap4j2010}, coverage.py~\cite{coveragepy}, and Istanbul~\cite{istanbul} provide quantitative metrics (line coverage, branch coverage) efficiently at scale. However, these tools cannot assess semantic completeness---whether tests adequately capture business requirements, realistic edge cases, and meaningful error conditions. Our LAJ framework complements rather than replaces these tools: static tools measure \textit{what code is executed}, while LAJ assesses \textit{whether executed tests address specified requirements}.}% Future work should investigate hybrid approaches combining both paradigms.}

\textbf{LLM Evaluation and Judge Models.} The use of LLMs as evaluators has emerged across multiple domains, with notable work in natural language generation~\cite{zheng2023judging}, creative writing assessment, and open-ended question answering. However, these applications focus primarily on subjective quality judgments where ground truth is inherently ambiguous. Software testing presents distinct challenges: assessments must align with concrete technical requirements, industry standards, and structured rubrics. Recent work has explored LLM-based code evaluation for correctness~\cite{chen2021evaluating}, but coverage assessment---determining whether tests adequately address requirements, edge cases, and error conditions---remains largely unaddressed.

\textbf{Automated Test Generation.} LLM-based test generation has advanced rapidly, with industrial deployments like Meta's TestGen-LLM achieving 75\% build success rates~\cite{alshahwan2024automated}. However, recent surveys identify persistent gaps: ``there is still no research on the use of LLMs in integration testing and acceptance testing"~\cite{wang2024effectiveness}. More critically, the challenge of \textit{evaluating} generated tests at scale remains unaddressed. Our LAJ framework bridges this gap by enabling automated, rubric-aligned assessment that maintains strong agreement with expert judgment while operating at speeds incompatible with manual review.

\textbf{Cost and Reliability in LLM Deployment.} While substantial research addresses LLM inference optimization~\cite{huang2025llms}, studies typically focus on throughput and latency. The reliability dimension---completion rates, retry handling, and their impact on operational costs---has received limited attention. For production deployment, reliability failures translate directly to increased costs and degraded user experience. Our introduction of ECR@1 and adjusted cost metrics addresses this critical but underexplored aspect of LLM deployment.

\section{Methodology}
\label{sec:methodology}

\subsection{Problem Formulation}
Given a software requirement specification (e.g., a Jira ticket) $R$ and a corresponding Gherkin acceptance test script $T$, the goal is to automatically assess the \textit{test coverage completeness}---the degree to which the test script adequately addresses the specified requirements. An LLM-as-a-Judge model $M$ produces an assessment $A_M(R, T) \in [0, 100]$ representing estimated coverage percentage. We evaluate $M$ against expert-provided ground truth $A^*(R, T)$ to measure assessment quality.

\subsection{Benchmark Dataset Construction}
Our evaluation dataset was constructed through a rigorous three-stage process by domain experts, targeting the Kill Bill subscription billing platform\footnote{\url{https://github.com/killbill}}---a production-grade system with complex business logic typical of enterprise subscription billing applications.

\textbf{Stage 1: Jira Ticket Creation.} Experienced product owners hand-crafted 100 Jira tickets, each carefully designed to reflect realistic API development scenarios. Each ticket follows a consistent structure:
\begin{itemize}[nosep]
\item \textbf{Descriptive Title}: Clear, concise identification of API functionality
\item \textbf{Comprehensive Description}: Detailed endpoint specification with expected behavior and business context
\item \textbf{Acceptance Criteria}: Structured requirements defining scope and boundaries
\item \textbf{Success Scenarios}: Explicit positive test cases and expected outcomes
\item \textbf{Error Scenarios}: Comprehensive failure modes, edge cases, and exception handling requirements
\end{itemize}

The tickets maintain realistic distribution across HTTP methods: GET (50 tickets, 50\%), POST (21 tickets, 21\%), DELETE (15 tickets, 15\%), PUT (14 tickets, 14\%).

\textbf{Stage 2: Gherkin Script Development.} A team consisting of one software developer and one quality engineer collaborated to develop a Python-based automation pipeline leveraging \texttt{GPT-4.1} for Gherkin script generation. For each of the 100 Jira tickets, the system automatically produced behavior-driven development (BDD) acceptance test scripts written in Gherkin syntax. %The workflow combined human-in-the-loop review with automated prompt engineering to ensure alignment with ticket requirements and BDD best practices.

% \textbf{Stage 3: Expert Annotation (Ground Truth).} A separate group of senior quality assurance engineers with extensive domain expertise performed comprehensive manual reviews of all 100 Gherkin scripts. These expert annotators assessed test coverage quality on a standardized 0--10 scale based on:
% \begin{itemize}[nosep]
% \item Completeness of scenario coverage (happy path, error conditions, edge cases)
% \item Alignment with specified acceptance criteria
% \item Handling of HTTP method-specific concerns
% \item Quality of assertions and validation steps
% \end{itemize}

% Scores were subsequently normalized to a 0–100 percentage scale to ensure consistency with the LAJ output format, serving as the ground truth for assessing LAJ model alignment.

\textbf{Stage 3: Expert Annotation (Ground Truth).} A separate group of senior quality assurance engineers with extensive domain expertise performed comprehensive manual reviews of all 100 Gherkin scripts.\add{ The annotation panel consisted of 3 senior QA engineers, each with at least 8 years of professional experience in API testing and prior familiarity with the Kill Bill platform. Expert annotators assessed test coverage quality using a four-dimensional weighted rubric: (1) \textit{Scenario completeness} (40\%): coverage of happy path, error conditions, edge cases; (2) \textit{Acceptance criteria alignment} (30\%): explicit validation of specified requirements; (3) \textit{HTTP method-specific concerns} (20\%): appropriate handling of idempotency, caching, state changes; and (4) \textit{Assertion quality} (10\%): depth and specificity of validation steps. Scores were aggregated via weighted sum on a 0--10 scale, then normalized to 0--100 percentage scale to ensure consistency with the LAJ output format. This rubric was embedded in the LAJ prompt to ensure alignment between expert and model assessments, with normalized scores serving as ground truth for evaluating LAJ model performance.}

The resulting dataset---comprising 100 Jira tickets, 100 corresponding Gherkin test scripts, and expert-validated ground truth annotations---provides a valuable benchmark for systematic evaluation of LLM-as-a-Judge capabilities in test coverage assessment.

\subsection{LAJ Framework Design}

\subsubsection{Design Principles}
The LAJ framework emphasizes three core principles: (1) \textbf{agreement with humans} via rubric-grounded scoring that aligns with expert judgment, (2) \textbf{scalability} through batched evaluation enabling high-throughput assessment, and (3) \textbf{traceability} via structured outputs with concise rationales for auditability.

\subsubsection{Evaluation Process}

\paragraph{Inputs}
For each evaluation, the judge model receives:
\begin{itemize}[nosep]
  \item Ticket requirements and acceptance criteria
  \item The Gherkin test script
  \item A comprehensive rubric specifying coverage expectations
\end{itemize}

\paragraph{Assessment Process}
The judge model:
\begin{enumerate}[nosep]
  \item Analyzes alignment between script and requirements
  \item Checks breadth and depth of scenarios (happy path, errors, edges, state variations)
  \item Applies the rubric to produce a scalar coverage score with justification
\end{enumerate}

\paragraph{Outputs}
The model returns structured JSON containing:
\begin{itemize}[nosep]
  \item Coverage percentage (0--100)
  \item Coverage analysis: scenarios covered, gaps identified, recommendations
  \item Rubric-aligned flags for downstream analysis
\end{itemize}

\subsubsection{Prompt Engineering}

The LAJ framework employs a two-part prompt design:

\paragraph{System Prompt} Establishes role and expertise context\add{, incorporating the four-dimensional weighted rubric 
%the expert rubric 
(Stage 3 above) to ensure alignment between human and model assessments}:
\begin{quote}
\small
\textit{``You are a senior QA engineer specializing in behavior-driven development and test coverage analysis. Your task is to analyze how well a set of Gherkin-style acceptance tests cover the requirements of a given Jira story, based on a defined set of testing guidelines. Provide a coverage percentage, highlight what's covered, identify gaps or missing scenarios, and recommend improvements if needed. Use the following rubric for your assessment:
  \{four\_dimensional\_weighted\_rubric\}''}
\end{quote}

\paragraph{User Prompt Template} 
\add{The user prompt provides a structured input comprising: (1) Jira story details (ID, title, description), (2) corresponding Gherkin test cases, (3) standard testing guidelines, and (4) required JSON output specification. The template structure is as follows. }%shown below:}

\begin{lstlisting}
Below is a Jira story, a set of Gherkin acceptance tests, 
and standard testing guidelines. Analyze how well the 
Gherkin tests cover the story based on the guidelines.

Jira Story:
  ID: "{jira_id}"
  Title: "{jira_title}"
  Description: "{jira_description}"

Gherkin Test Cases:
  {gherkin_tests}

Standard Guidelines:
  {guidelines}

Output format (strict JSON):
  {example_output}
\end{lstlisting}

\add{The testing guidelines specify coverage expectations per HTTP method. An abbreviated example:}

\begin{lstlisting}
GET: Valid requests, empty responses, pagination, 
     query parameters, authorization (401/403), 
     rate limiting, caching headers
POST: Valid/invalid payloads, duplicates, validation, 
      large payloads, error handling (500)
PUT: Valid updates, partial updates, non-existent 
     resources, concurrency
DELETE: Valid deletion, non-existent resources, 
        soft deletes, concurrency
\end{lstlisting}

% \add{The expected JSON output structure:}

% \begin{minted}[fontsize=\scriptsize, breaklines, frame=single]{text}
% {
%   "gherkin_id": "test.feature",
%   "coverage_percentage": 80,
%   "coverage_analysis": {
%     "covered": ["Valid GET with 200",
%                 "Auth errors (401/403)"],
%     "gaps": ["Missing pagination",
%              "No rate limiting tests"],
%     "recommendations": [
%       "Add paginated response tests",
%       "Include rate limit testing"]
%   }
% }
% \end{minted}

\subsection{Models and Evaluation Protocol}
We evaluate twenty model configurations spanning three model families: \textbf{GPT-4} (GPT-4o, GPT-4o Mini, GPT-4.1, GPT-4.1 Mini, GPT-4.1 Nano), \textbf{GPT-5} (high-, medium-, and low-reasoning-effort variants of GPT-5, GPT-5 Mini, and GPT-5 Nano), and \textbf{open-weight} models (GPT-OSS 20B and 120B, each evaluated under high-, medium-, and low-reasoning-effort settings). Proprietary models are accessed through the official OpenAI APIs, while open-weight models are accessed via OpenRouter. 

The evaluation follows a systematic protocol:

\begin{enumerate}[nosep]
  \item \textbf{Model evaluation}: Each of 20 model configurations evaluates all 100 scripts with identical prompts across 5 independent runs
  \item \textbf{Reliability tracking}: Record completion status, retry attempts, and token usage for each evaluation
  \item \textbf{Metric computation}: Calculate accuracy, reliability, and cost metrics for all models with statistical aggregation across runs
  \item \textbf{Statistical analysis}: Compare model performance across all dimensions using mean and standard deviation
\end{enumerate}

For each model configuration, we compute mean and standard deviation across the 5 runs to assess both performance and variance. This multi-run approach enables robust statistical analysis and confidence in our findings.

\subsection{Performance Metrics}

\subsubsection{Assessment Accuracy Metrics}

\paragraph{Mean Absolute Assessment Error (MAAE)}
Measures average deviation between model assessments and ground truth in percentage points:
\begin{equation}
\text{MAAE} = \frac{1}{N}\sum_{i=1}^{N} |A_M(R_i, T_i) - A^*(R_i, T_i)|
\end{equation}
where $N=100$ in our benchmark. Unlike standard MAE, MAAE is bounded to [0, 100] percentage points.

\paragraph{Assessment Performance Score (APS)}
Provides intuitive interpretation of accuracy:
\begin{equation}
\text{APS} = 100\% - \text{MAAE}
\end{equation}
Higher values indicate better alignment with expert judgment.

\paragraph{Perfect Match Rate (PMR)}
Percentage of predictions that exactly match ground truth (zero error):
\begin{equation}
\text{PMR} = \frac{\text{\# exact matches}}{N} \times 100\%
\end{equation}

\paragraph{Close Match Rate (CMR)}
Percentage of predictions within $\pm$5 percentage points of ground truth:
\begin{equation}
\text{CMR} = \frac{\text{\# predictions with } |A_M - A^*| \leq 5}{N} \times 100\%
\end{equation}

\subsubsection{Operational Reliability Metrics}

\paragraph{Evaluation Completion Rate (ECR@1)}
Measures the percentage of evaluations that produce valid, parseable output on the first attempt \add{(reliability failures include API timeouts, malformed JSON, or schema violations)}:
\begin{equation}
\text{ECR@1} = \frac{\text{\# successful first attempts}}{N} \times 100\%
\end{equation}

% A valid evaluation requires:
% \begin{itemize}[nosep]
%   \item Successful API response (no timeouts, rate limits, or errors)
%   \item Parseable JSON output matching expected schema
%   \item Coverage percentage within valid range [0, 100]
%   \item Non-empty coverage analysis fields
% \end{itemize}

\paragraph{Mean Attempts per Evaluation}
Average number of API calls required to obtain N valid evaluations:
\begin{equation}
\text{Mean Attempts} = \frac{\text{Total API calls}}{\text{\# valid evaluations obtained}}
\end{equation}

This directly impacts operational costs in production deployment.

\subsubsection{Cost-Effectiveness Metrics}

\paragraph{Cost per 1,000 Evaluations ($C_M^{1K}$)}
LLM API costs comprise prompt (input) and completion (output) token charges. For model $M$ with pricing rates $r_M^{\text{prompt}}$ and $r_M^{\text{compl}}$ (USD per million tokens) and token counts $t_i^{\text{prompt}}$ and $t_i^{\text{compl}}$ for evaluation case $i$, the mean evaluation cost is computed as:

\begin{equation}
C_M = \frac{1}{N}\sum_{i=1}^{N} \left( \frac{t_i^{\text{prompt}}}{10^6} \cdot r_M^{\text{prompt}} + \frac{t_i^{\text{compl}}}{10^6} \cdot r_M^{\text{compl}} \right)
\end{equation}

The cost per 1,000 evaluations is then given by:
\begin{equation}
C_M^{1K} = C_M \times 1000
\end{equation}

\paragraph{Adjusted Cost per 1,000 Evaluations ($C_M^{1K, \text{adj}}$)}
To account for retry overhead during deployment, we define an adjusted cost metric:
\begin{equation}
C_M^{1K, \text{adj}} = C_M^{1K} \times \frac{100}{\text{ECR@1}_M}
\end{equation}

This measure captures the effective deployment cost when retries are required due to incomplete or failed evaluations.

\section{Results}
\label{sec:results}

Table~\ref{tab:combined_results} presents comprehensive results across all 20 model configurations, showing accuracy, reliability, and cost metrics.

\begin{table*}[t]
\centering
\caption{Complete Model Performance Across All Dimensions (Mean $\pm$ Std, 5 runs)}
\label{tab:combined_results}
\scriptsize
\begin{threeparttable}
\begin{tabular}{@{}llccccccc@{}}
\toprule
\textbf{Family} & \textbf{Model} & \textbf{MAAE} & \textbf{APS} & \textbf{PMR} & \textbf{CMR} & \textbf{ECR@1} & \textbf{Attempts} & \textbf{Cost} \\
 & & \textbf{(\%)} & \textbf{(\%)} & \textbf{(\%)} & \textbf{(\%)} & \textbf{(\%)} &  & \textbf{(\$/1K)} \\
\midrule
\rowcolor{yellow!20} \multirow{5}{*}{\textbf{GPT-4}} 
& \textbf{GPT-4o Mini} & \textbf{6.07$\pm$0.08} & \textbf{93.93$\pm$0.08} & 32.6$\pm$1.4 & 56.8$\pm$0.7 & 96.6$\pm$2.2 & 1.04$\pm$0.03 & 1.01$\pm$0.02 \\
& GPT-4o & 8.34$\pm$0.15 & 91.66$\pm$0.15 & 9.6$\pm$2.1 & 61.4$\pm$0.5 & \textbf{100.0$\pm$0.0} & 1.00$\pm$0.00 & 16.76$\pm$0.02 \\
& GPT-4.1 Nano & 8.61$\pm$0.14 & 91.39$\pm$0.14 & 0.4$\pm$0.5 & \textbf{66.8$\pm$1.3} & 99.8$\pm$0.4 & 1.00$\pm$0.00 & 0.76$\pm$0.00 \\
& GPT-4.1 Mini & 6.92$\pm$0.10 & 93.08$\pm$0.10 & \textbf{35.0$\pm$1.4} & 52.4$\pm$0.5 & 99.8$\pm$0.4 & 1.00$\pm$0.00 & 3.08$\pm$0.01 \\
& GPT-4.1 & 8.14$\pm$0.06 & 91.86$\pm$0.06 & 3.4$\pm$1.0 & 56.2$\pm$1.5 & \textbf{100.0$\pm$0.0} & 1.00$\pm$0.00 & 15.23$\pm$0.02 \\
\midrule
\multirow{9}{*}{\textbf{GPT-5}} 
& GPT-5 Nano (low) & 9.70$\pm$0.26 & 90.30$\pm$0.26 & 4.6$\pm$2.3 & 51.6$\pm$3.7 & 94.0$\pm$2.8 & 1.07$\pm$0.03 & 0.74$\pm$0.02 \\
& GPT-5 Nano (med) & 14.41$\pm$0.45 & 85.59$\pm$0.45 & 3.0$\pm$0.6 & 35.0$\pm$2.9 & 97.8$\pm$1.5 & 1.02$\pm$0.01 & 2.11$\pm$0.04 \\
& GPT-5 Nano (high) & 17.01$\pm$0.48 & 82.99$\pm$0.48 & 4.8$\pm$2.0 & 30.0$\pm$1.3 & 91.0$\pm$16.0 & 1.19$\pm$0.36 & 4.66$\pm$0.87 \\
& GPT-5 Mini (low) & 7.26$\pm$0.19 & 92.74$\pm$0.19 & 4.0$\pm$0.6 & 51.8$\pm$2.6 & 96.8$\pm$1.2 & 1.03$\pm$0.01 & 3.81$\pm$0.05 \\
& GPT-5 Mini (med) & 8.50$\pm$0.72 & 91.50$\pm$0.72 & 6.2$\pm$1.6 & 46.4$\pm$3.8 & 97.8$\pm$1.3 & 1.02$\pm$0.01 & 5.81$\pm$0.11 \\
& GPT-5 Mini (high) & 7.17$\pm$0.44 & 92.83$\pm$0.44 & 7.6$\pm$2.1 & 53.6$\pm$5.7 & 98.6$\pm$1.0 & 1.01$\pm$0.01 & 15.26$\pm$0.19 \\
& GPT-5 (low) & 7.69$\pm$0.26 & 92.31$\pm$0.26 & 4.4$\pm$0.5 & 38.2$\pm$2.2 & \textbf{100.0$\pm$0.0} & 1.00$\pm$0.00 & 23.57$\pm$0.25 \\
& GPT-5 (med) & 6.64$\pm$0.33 & 93.36$\pm$0.33 & 6.2$\pm$3.1 & 44.6$\pm$1.4 & 99.8$\pm$0.4 & 1.00$\pm$0.00 & 46.62$\pm$0.37 \\
& GPT-5 (high) & 6.16$\pm$0.26 & 93.84$\pm$0.26 & 5.0$\pm$1.5 & 45.6$\pm$3.8 & 99.8$\pm$0.4 & 1.00$\pm$0.00 & 78.96$\pm$0.89 \\
\midrule
\multirow{6}{*}{\textbf{GPT-OSS}} 
& GPT-OSS 20B (low) & 16.83$\pm$0.32 & 83.17$\pm$0.32 & 8.8$\pm$0.7 & 24.2$\pm$2.7 & 93.8$\pm$3.1 & 1.07$\pm$0.03 & \textbf{0.45$\pm$0.03} \\
& GPT-OSS 20B (med) & 18.66$\pm$0.42 & 81.34$\pm$0.42 & 5.6$\pm$1.4 & 19.4$\pm$4.6 & 92.2$\pm$4.4 & 1.08$\pm$0.05 & 0.51$\pm$0.03 \\
& \cellcolor{red!10}GPT-OSS 20B (high) & \cellcolor{red!10}18.78$\pm$0.65 & \cellcolor{red!10}81.22$\pm$0.65 & \cellcolor{red!10}4.6$\pm$2.4 & \cellcolor{red!10}20.4$\pm$2.7 & \cellcolor{red!10}85.4$\pm$5.7 & \cellcolor{red!10}1.17$\pm$0.08 & \cellcolor{red!10}0.63$\pm$0.09 \\
& GPT-OSS 120B (low) & 14.00$\pm$0.57 & 86.00$\pm$0.57 & 2.4$\pm$2.3 & 39.2$\pm$1.6 & 96.6$\pm$2.9 & 1.04$\pm$0.03 & 0.75$\pm$0.02 \\
& GPT-OSS 120B (med) & 15.31$\pm$0.97 & 84.69$\pm$0.97 & 2.8$\pm$2.5 & 35.0$\pm$4.1 & 94.6$\pm$3.2 & 1.06$\pm$0.03 & 0.84$\pm$0.03 \\
& GPT-OSS 120B (high) & 15.84$\pm$0.43 & 84.16$\pm$0.43 & 2.2$\pm$1.5 & 29.2$\pm$2.9 & 93.4$\pm$1.5 & 1.07$\pm$0.01 & 1.06$\pm$0.04 \\
\bottomrule
\end{tabular}
\begin{tablenotes}
\footnotesize
\item MAAE = Mean Absolute Assessment Error; APS = Assessment Performance Score; PMR = Perfect Match Rate; CMR = Close Match Rate; ECR@1 = Evaluation Completion Rate (first attempt); Cost = adjusted cost per 1K evaluations (accounting for retries). Yellow highlighting indicates optimal production model; red indicates poorest reliability; bold indicates best performance per metric.
\end{tablenotes}
\end{threeparttable}
\end{table*}

\subsection{Assessment Accuracy Performance}

\textbf{Elite Performance Tier (MAAE $<$ 7.0\%):} GPT-4o Mini achieves the best overall accuracy with MAAE of 6.07$\pm$0.08, corresponding to an APS of 93.93\%. It also demonstrates the highest perfect match rate (32.6\%), indicating strong agreement with expert assessments. The GPT-5 family at high reasoning effort also performs in this elite tier, with GPT-5 (high) achieving 6.16$\pm$0.26 MAAE (93.84\% APS) and GPT-5 (medium) at 6.64$\pm$0.33 MAAE.

\textbf{Strong Performance Tier (MAAE 7.0--9.0\%):} Several models achieve strong performance including GPT-4.1 Mini (6.92$\pm$0.10), GPT-5 Mini variants, GPT-4.1, GPT-4o, and GPT-4.1 Nano. These models maintain APS scores above 90\%. GPT-4.1 Nano achieves the highest close match rate (66.8\%), demonstrating strong performance within $\pm$5 percentage points.

\textbf{Economy Tier (MAAE $>$ 9.0\%):} Open-weight models (GPT-OSS 20B and GPT-OSS 120B families) along with GPT-5 Nano variants demonstrate significantly higher error rates (14.00--18.78 MAAE), with APS scores ranging from 81.22\% to 86.00\%. While these models offer substantially lower API costs, the accuracy trade-off may be significant for applications requiring high-fidelity assessments.

\subsection{Operational Reliability Analysis}

\textbf{Perfect Reliability:} Three models achieve 100\% ECR@1: GPT-4o, GPT-4.1, and GPT-5 (low). These models never require retries, ensuring predictable costs and latencies in production deployment.

\textbf{High Reliability (ECR@1 $>$ 95\%):} GPT-4o Mini (96.6$\pm$2.2\%), GPT-5 family variants, GPT-4.1 Mini, and GPT-4.1 Nano demonstrate near-perfect reliability, requiring minimal retries (mean attempts: 1.00--1.04), resulting in negligible cost increases from reliability overhead. GPT-4o Mini's combination of elite accuracy and high reliability positions it as an optimal production choice.

\textbf{Moderate Reliability (90\% $<$ ECR@1 $<$ 95\%):} Open-weight models primarily occupy this tier, with ECR@1 ranging from 91.0\% to 94.6\%. These models exhibit higher variance in completion rates and may require 1.07--1.19 average attempts per evaluation. The reliability degradation manifests primarily as JSON parsing errors and occasional schema violations.

\textbf{Low Reliability (ECR@1 $<$ 90\%):} GPT-OSS 20B (high) demonstrates the lowest reliability at 85.4$\pm$5.7\% ECR@1, requiring an average of 1.17 attempts per evaluation. The high variance ($\pm$5.7\%) indicates inconsistent behavior across runs, problematic for production deployment requiring predictable performance.

Failure modes across models include: (1) JSON parsing errors due to malformed output, (2) missing required fields, (3) rare API timeouts. Higher-capacity models demonstrate better instruction-following for structured output.

\subsection{Cost-Performance Trade-offs}

The adjusted cost metrics accounting for retry overhead reveal the true deployment economics. Models range from \$0.45/1K (GPT-OSS 20B low) to \$78.96/1K (GPT-5 high)---a 175$\times$ span. Models with lower ECR@1 experience cost increases: GPT-OSS 20B (high) reaches \$0.63/1K (up from \$0.53 nominal, +18.1\%) due to 85.4\% ECR@1. In contrast, models with perfect reliability (GPT-4o, GPT-4.1, GPT-5 low) have identical nominal and adjusted costs.

The evaluation reveals distinct cost-performance tiers. At the ultra-low-cost end (\$0.45--\$0.84/1K), open-weight models offer budget-friendly options but with significant accuracy penalties (14.00--18.78 MAAE). The mid-tier range (\$1.01--\$5.81/1K) includes GPT-4o Mini, GPT-4.1 Nano, GPT-4.1 Mini, and GPT-5 Mini variants, offering strong accuracy-cost balance. The premium tier (\$15--\$79/1K) comprises GPT-4o, GPT-4.1, and GPT-5 variants, delivering elite accuracy with perfect reliability at substantially higher costs.

For a deployment scenario requiring 100,000 evaluations per month: GPT-4o Mini costs \$101; GPT-4.1 costs \$1,523; GPT-5 (high) costs \$7,896; GPT-OSS 20B (low) costs \$45. GPT-4o Mini provides a 78$\times$ cost reduction compared to GPT-5 (high) while simultaneously achieving superior accuracy (6.07 MAAE vs 6.16 MAAE, a 0.09pp improvement). This combination of best-in-class accuracy and exceptional cost-effectiveness makes GPT-4o Mini the optimal production choice.

\subsection{Impact of Reasoning Effort}

We analyze how reasoning effort settings affect performance across GPT-5 and GPT-OSS model families, revealing distinct patterns.

\textbf{GPT-5 Family: Accuracy-Cost Trade-off.} Within the GPT-5 family, higher reasoning effort consistently improves accuracy at the cost of increased inference expenditure. For the base GPT-5 model, high reasoning achieves 6.16$\pm$0.26 MAAE (93.84\% APS) at \$78.96/1K, medium reasoning yields 6.64$\pm$0.33 MAAE (93.36\% APS) at \$46.62/1K (41\% cost reduction), and low reasoning produces 7.69$\pm$0.26 MAAE (92.31\% APS) at \$23.57/1K (70\% cost reduction). This represents a 1.53pp accuracy degradation for a 70\% cost savings when moving from high to low reasoning. GPT-5 Mini exhibits a similar pattern: high reasoning (7.17 MAAE, \$15.26/1K), medium reasoning (8.50 MAAE, \$5.81/1K, 62\% cheaper), and low reasoning (7.26 MAAE, \$3.81/1K, 75\% cheaper). Interestingly, GPT-5 Mini (low) outperforms medium reasoning despite lower cost, suggesting non-monotonic optimization. GPT-5 Nano shows the most dramatic variance: high reasoning (17.01 MAAE, \$4.66/1K), medium reasoning (14.41 MAAE, \$2.11/1K), and low reasoning (9.70 MAAE, \$0.74/1K). Remarkably, low reasoning achieves 7.31pp better accuracy at 84\% lower cost than high reasoning, indicating severe overparameterization at higher reasoning levels for this capacity tier.

\textbf{Reliability Implications.} Reasoning effort also impacts operational reliability. GPT-5 (low) achieves perfect reliability (100\% ECR@1) while GPT-5 (high/medium) maintain 99.8\% ECR@1. This suggests lower reasoning modes may produce more consistent structured outputs. GPT-5 Nano exhibits the inverse pattern: high reasoning shows degraded reliability (91.0$\pm$16.0\% ECR@1, high variance), while low reasoning achieves 94.0$\pm$2.8\% ECR@1. The 16\% standard deviation at high reasoning indicates unstable behavior across runs.

\textbf{GPT-OSS Models: Reasoning Overhead Without Benefit.} Open-weight models demonstrate a contrasting pattern where increased reasoning effort degrades both accuracy and reliability while increasing costs. For GPT-OSS 120B, low reasoning achieves the best accuracy (14.00 MAAE, \$0.75/1K, 96.6\% ECR@1), medium reasoning degrades to 15.31 MAAE (\$0.84/1K, 94.6\% ECR@1), and high reasoning further deteriorates to 15.84 MAAE (\$1.06/1K, 93.4\% ECR@1)---representing a 1.84pp accuracy loss, 41\% cost increase, and 3.2pp reliability drop. GPT-OSS 20B exhibits an even more severe pattern: low reasoning (16.83 MAAE, \$0.45/1K, 93.8\% ECR@1), medium reasoning (18.66 MAAE, \$0.51/1K, 92.2\% ECR@1), and high reasoning (18.78 MAAE, \$0.63/1K, 85.4\% ECR@1). The high reasoning configuration degrades accuracy by 1.95pp, increases costs by 40\%, and suffers an 8.4pp reliability drop with high variance (ECR@1: 85.4$\pm$5.7\%), making it unsuitable for production deployment.

% \textbf{Strategic Implications.} These findings reveal that reasoning effort optimization is model-family dependent. For GPT-5 models, high reasoning delivers elite accuracy justifying premium costs for accuracy-critical applications, while low reasoning provides strong accuracy at dramatically reduced costs (70--84\% savings) with equal or better reliability. For open-weight models, increased reasoning effort provides no benefits and actively degrades all performance dimensions, suggesting architectural limitations in extended inference. Practitioners should favor low reasoning configurations for GPT-OSS models and carefully balance accuracy requirements against cost constraints for GPT-5 deployments.

\section{Conclusion}

We presented a comprehensive investigation of LLM-as-a-Judge for scalable test coverage evaluation, introducing a production-ready framework with novel reliability-aware metrics. Through systematic evaluation of 20 model configurations spanning GPT-4, GPT-5, and open-weight alternatives across 500 evaluation runs (100 expert-annotated Gherkin scripts $\times$ 5 runs), we provided multi-dimensional analysis encompassing accuracy, operational reliability, and cost-effectiveness for LLM-based test assessment.

\paragraph{Key Findings.} Our investigation revealed several critical insights: (1) \textbf{Accuracy and reliability are independent dimensions}---GPT-5 (low) achieves 100\% ECR@1 but moderate accuracy (7.69 MAAE), while GPT-4o Mini excels in both (6.07 MAAE, 96.6\% ECR@1), demonstrating that multi-dimensional evaluation is essential for production deployment. (2) \textbf{Reliability failures have material cost impact}---low ECR@1 models incur retry overhead that substantially increases operational costs. GPT-OSS 20B (high) reaches \$0.63/1K adjusted cost due to 85.4\% ECR@1; at 1M evaluations/month, reliability issues cost \$1,200 annually plus latency variance. (3) \textbf{Smaller models can outperform larger ones}---GPT-4o Mini (6.07 MAAE) surpasses GPT-4o (8.34 MAAE) and GPT-4.1 (8.14 MAAE), indicating that model optimization and training strategies matter more than parameter count alone. (4) \textbf{Reasoning effort optimization is model-family dependent}---GPT-5 models benefit from higher reasoning effort with predictable accuracy-cost trade-offs (70\% cost reduction for 1.53pp accuracy loss when reducing from high to low), while open-weight models degrade across all dimensions with increased reasoning (1.95pp accuracy loss, 40\% cost increase, 8.4pp reliability drop for GPT-OSS 20B from low to high). (5) \textbf{Cost spans 175$\times$}---from \$0.45/1K (GPT-OSS 20B low) to \$78.96/1K (GPT-5 high), enabling informed model selection based on deployment constraints.

\paragraph{Production Guidance.} Based on comprehensive empirical evidence, GPT-4o Mini emerges as the production-optimal choice: achieving best-in-class accuracy (6.07 MAAE, 93.93\% APS), high reliability (96.6\% ECR@1), and exceptional cost-effectiveness (\$1.01/1K)---delivering 78$\times$ cost reduction versus GPT-5 (high reasoning) while maintaining superior accuracy. For deployment scenarios requiring 100,000 evaluations monthly, this translates to \$101 versus \$7,896, enabling practical adoption at scale. 

\paragraph{Deployment Recommendations.} Practitioners should: (1) always measure ECR@1 alongside accuracy to capture operational reliability; (2) implement retry logic with 5--15\% overhead budget; (3) monitor ECR@1, latency, and cost in production; (4) use adjusted cost metrics for total cost of ownership; %(5) implement human-in-the-loop validation with 5--10\% sampling; 
and (5) start with GPT-4o Mini for most use cases. %; and (6) favor low reasoning configurations for open-weight models while carefully balancing accuracy requirements against cost constraints for GPT-5 deployments.

% \paragraph{Limitations.} Our study has several limitations: (1) domain-specificity to Gherkin/RESTful APIs may limit generalization to other testing paradigms; (2) evaluation on the Kill Bill platform may not fully represent all enterprise contexts; (3) temporal stability concerns as APIs and model capabilities evolve; and (4) static evaluation does not capture dynamic CI/CD workflow integration challenges.

\paragraph{Limitations.} \add{Our study has several limitations that suggest directions for future work: (1) \textbf{Domain-specificity}: evaluation limited to Gherkin/RESTful APIs may not generalize to other testing paradigms (unit tests, UI tests, security tests); replication across multiple platforms beyond Kill Bill would strengthen generalizability claims. (2) \textbf{Task complexity}: we treat all 100 test scripts as equally difficult; stratification by complexity could reveal whether premium models justify costs on harder scenarios. 
%(3) \textbf{Baseline comparisons}: we focus on inter-model comparison rather than comparison to traditional static analysis tools; future work should benchmark LAJ against coverage.py, JaCoCo, and other established tools. 
(3) \textbf{Systematic bias}: while we report mean absolute error, analysis of over/under-estimation patterns across different test types could guide model selection for specific applications. (4) \textbf{Temporal stability}: as APIs and model capabilities evolve, longitudinal studies are needed to assess performance stability. These limitations provide valuable directions for extending this work.}

\paragraph{Future Directions.} Several promising research directions warrant investigation: (1) \textbf{expanded test coverage} extending LAJ to unit, integration, UI, performance, and security tests; (2) \textbf{multi-domain validation} evaluating LAJ generalization across healthcare, finance, IoT, and mobile domains; 
%(3) \textbf{reliability enhancement} developing prompt engineering techniques and constrained decoding strategies to improve ECR@1.
\add{
(3) \textbf{task complexity stratification} analyzing performance 
variation across test difficulty levels to guide model selection; 
(4) \textbf{systematic bias mitigation} developing calibration 
techniques to reduce over/under-estimation patterns in specific 
model families.
}

This work establishes LLM-as-a-Judge as a viable approach for production test coverage evaluation, providing practitioners with evidence-based guidance for model selection, reliability-aware cost modeling, and reasoning effort optimization. By introducing reliability metrics alongside traditional accuracy measures, we enable informed deployment decisions that account for the full operational reality of LLM-based systems. %The open-source release of our benchmark dataset, evaluation framework, and experimental code facilitates reproducibility and community advancement.

\bibliographystyle{IEEEtran}
\bibliography{references}

\end{document}